\documentclass[aps,prl,superscriptaddress,twocolumn]{revtex4}

\usepackage{amsmath,amssymb}
\usepackage{graphicx}
\usepackage{longtable}
\usepackage[ansinew]{inputenc}
\usepackage{color}

\newcommand{\beq}{\begin{eqnarray}}
\newcommand{\eeq}{\end{eqnarray}}

\renewcommand{\Re}{{\rm Re}\,}
\renewcommand{\Im}{{\rm Im}\,}

\begin{document}


\title{In-plane magnetic anisotropy of Fe atoms on Bi$_2$Se$_3$(111)}

\author{J. Honolka}
 \email{j.honolka@fkf.mpg.de}
 \affiliation{Max-Planck-Institut f\"ur Festk\"orperforschung, Heisenbergstrasse 1, 70569 Stuttgart, Germany}

\author{A. A. Khajetoorians}
 \email[Corresponding author: ]{akhajeto@physnet.uni-hamburg.de}
\affiliation{Institute for Applied Physics, Universit\"{a}t Hamburg, D-20355 Hamburg, Germany}

\author{V. Sessi}
\affiliation{ESRF, Grenoble, France}

\author{T. O. Wehling}
\affiliation{I. Institut f\"{u}r Theoretische Physik, Universit\"{a}t Hamburg, D-20355 Hamburg, Germany}

\author{S. Stepanow}
 \affiliation{Max-Planck-Institut f\"ur Festk\"orperforschung, Heisenbergstrasse 1, 70569 Stuttgart, Germany}

\author{J.-L. Mi}
\affiliation{Center for Materials Crystallography, Department of Chemistry, Interdisciplinary Nanoscience Center, Aarhus University, 8000 Aarhus C, Denmark}

\author{B. B. Iversen}
\affiliation{Center for Materials Crystallography, Department of Chemistry, Interdisciplinary Nanoscience Center, Aarhus University, 8000 Aarhus C, Denmark}

\author{T. Schlenk}
\affiliation{Institute for Applied Physics, Universit\"{a}t Hamburg, D-20355 Hamburg, Germany}

\author{J. Wiebe}
\affiliation{Institute for Applied Physics, Universit\"{a}t Hamburg, D-20355 Hamburg, Germany}

\author{N. Brookes}
\affiliation{ESRF, Grenoble, France}

\author{A. I. Lichtenstein}
\affiliation{I. Institut f\"{u}r Theoretische Physik, Universit\"{a}t Hamburg, D-20355 Hamburg, Germany}

\author{Ph. Hofmann}
\affiliation{Department of Physics and Astronomy, Interdisciplinary Nanoscience Center, Aarhus University, 8000 Aarhus C, Denmark}

\author{K. Kern}
\affiliation{Max-Planck-Institut f\"ur Festk\"orperforschung, Heisenbergstrasse 1, 70569 Stuttgart, Germany}
\affiliation{Institut de Physique de la Mati\`ere Condens\'ee, Ecole Polytechnique F\'ed\'erale de Lausanne, CH-1015 Lausanne, Switzerland}

\author{R. Wiesendanger}
\affiliation{Institute for Applied Physics, Universit\"{a}t Hamburg, D-20355 Hamburg, Germany}

\date{\today}

\begin{abstract}
\noindent

The robustness of the gapless topological surface state hosted by a 3D topological insulator against perturbations of magnetic origin has been the focus of recent investigations. We present a comprehensive study of the magnetic properties of Fe impurities on a prototypical 3D topological insulator Bi$_2$Se$_3$ using local low temperature scanning tunneling microscopy and integral x-ray magnetic circular dichroism techniques. Single Fe adatoms on the Bi$_2$Se$_3$ surface, in the coverage range $\approx 1\%$ are heavily relaxed into the surface and exhibit a magnetic easy axis within the surface-plane, contrary to what was assumed in recent investigations on the opening of a gap. Using \textit{ab initio} approaches, we demonstrate that an in-plane easy axis arises from the combination of  the crystal field and dynamic hybridization effects.

\end{abstract}

\pacs{}

\keywords{topological insulators}

\maketitle

Topological insulators (TI) have demanded heavy  interest from the scientific community as a new class of materials illuminating fascinating yet exotic physics and offering large potential for applications in the field of spintronics~\cite{HasanOverview}. TI host a gapless topological surface state (TSS) which exhibits a Dirac-cone like dispersion. However, unlike in the case of graphene, the Dirac cone is located in the center of the Brillouin-zone and the spin and momentum degrees of freedom are locked. The latter so-called topological quality of the TSS is protected by time-reversal symmetry which leads to a variety of interesting effects. For example, these materials may be a forum for Majorana fermions~\cite{Fu2009}, a topological magnetoelectric effect~\cite{Qi2008}, and a quantized anomalous Hall effect~\cite{Yu2010}.

The locking of both spin and momentum degrees of freedom leads to the suppression of $180^{\circ}$ elastic back-scattering of TSS electrons in the absence of spin-flip processes.  The robustness of these "topologically protected" processes, when introducing impurities which break time reversal symmetry, is of critical importance for spin-based transport in such materials.  It has been suggested that the interaction between magnetic impurities and the topological state can cause an opening of an energy gap at the Dirac point (DP), provided that the magnetic order is oriented normal to the surface plane~\cite{Hor2010, Chen2010, Wray2011}. Nevertheless, the stability of the TSS against local magnetic perturbations is currently under heavy debate~\cite{Biswas2010, arxivpaperScholz,Hofmann2011}, since little is known about the fundamental interface effects of magnetic entities with TI surfaces as well as the static magnetic properties of impurities in these materials.

In this work we present a comprehensive study of the magnetic properties of iron impurities on a prototypical TI Bi$_2$Se$_3$ using local scanning tunneling microscopy (STM) and integral x-ray magnetic circular dichroism (XMCD) techniques. We show that the multiplet structure visible in x-ray absorption spectra reflect a high-spin state of the adsorbed Fe impurities, which are confirmed to bind at hollow sites of the surface lattice, under the influence of trigonal crystal fields given in hollow site positions of the Bi$_2$Se$_3$ surface. In contrary to recent reports a significant in-plane magnetic easy axis direction is observed in the XMCD. This observation is supported by our DFT calculations which reveal the possibility of a preferred in-plane easy axis when considering both crystal field and dynamic hybridization effects.

Both local scanning tunneling spectroscopy (STS) and integral XMCD measurements were carried out with one and the same Bi$_2$Se$_3$
crystal used in Ref.~\cite{Bianchi2010}. Clean Bi$_2$Se$_3$(111) surfaces were obtained by {\it in situ} cleaving at room temperature under ultra-high
vacuum conditions followed by a fast cool-down to temperatures of about $T=10$~K within 15 minutes. In order to
prevent surface diffusion, Fe was deposited at lowest temperatures onto Bi$_2$Se$_3$(111) using an e-beam evaporator.
For the case of x-ray measurements the general procedure of {\it in situ} metal atom deposition and coverage
calibration is described in detail in Ref.~\cite{Honolka1}.

\begin{figure}
\includegraphics[width=\columnwidth]{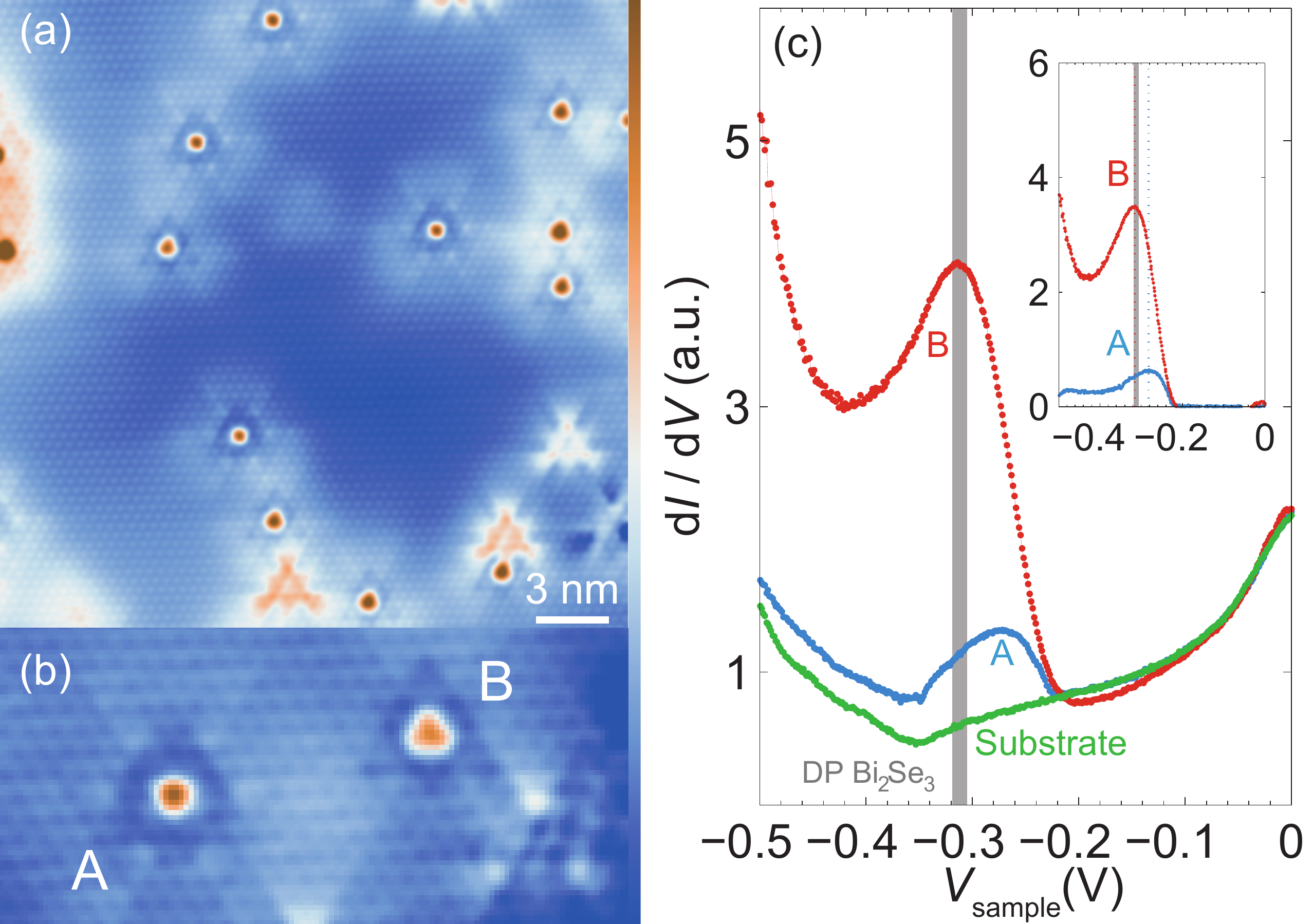}
\caption{\label{STM} (a) STM constant-current image of $1\%$ML of Fe atoms adsorbed onto a cleaved Bi$_2$Se$_3$ surface; $I_{\rm stab}=500$~pA, $V_{\rm stab}=+100$~mV. (b) High resolution image of two Fe atoms at different binding sites (A,B); $I_{\rm stab}=600$~pA, $V_{\rm stab}=-100$~mV. (c) Tunneling spectra taken on two Fe atoms at the indicated binding sites and on a substrate spot of the Fe covered surface. The gray bar refers to the DP of the uncovered Bi$_{2}$Se$_{3}$ surface as confirmed by probing the zeroth Landau level in high magnetic fields. Inset: same as (c) after subtracting the substrate spectrum; $I_{\rm stab}=600$~pA, $V_{\rm stab}=-100$~mV, $V_{\rm mod}=3$~mV; $f_{\rm mod} = 4.1$~kHz.}
\end{figure}

STS was performed at a temperature of $T=0.3$~K using a home-built ultra-high vacuum STM with a tungsten tip~\cite{Wiebe2004a}. Figure~\ref{STM}(a) shows $\approx1\%$ of a monolayer (ML) Fe deposited on a clean Bi$_2$Se$_3$ surface imaged with constant-current mode. Single Fe atoms are located with an almost equal ratio in two different hollow site positions with respect to the underlying hexagonal lattice of Se surface atoms. High resolution images (Fig.~\ref{STM}(b)) show that the atoms on the two binding sites, labeled A and B, clearly differ in apparent height and shape. The apparent height of 0.5~\AA~ and 0.3~\AA~for A and B, respectively, is small compared to e.g. Fe on Cu(111) ($\approx 1$~\AA, ~\cite{Khajetoorians2011}).
A closer inspection reveals that the apparent shape of the Fe and of the halo (charge density) surrounding the atom depends on the binding site. Type A atoms show a circular protrusion around the center of the Fe atom while type B atoms show a triangular protrusion. The surrounded halo pattern is quite complex but unique for each binding site. Infrequent hopping events of a single Fe atom between A and B stacking positions indicate a transformation of atom shape and halo between the two distinguishable atom types and binding sites. This proves that these properties are not due to underlying surface defects but are purely electronic effects resulting from the bonding between the Fe atom and the atomically flat Bi$_2$Se$_3$ surface.

Tunneling spectra (${\rm d}I/{\rm d}V(V)$) that were taken in different magnetic fields perpendicular to the surface on the bare Bi$_2$Se$_3$ surface prior to Fe deposition (not shown) exhibit Landau levels (LLs) with the expected energy scaling confirming the Dirac cone dispersion of the topological surface state~\cite{Hanaguri,Xue2010} as well as the position of the DP which is measured in a window  around $E_\mathrm{DP}$ = -0.31 eV. This is consistent with ARPES measurements of the band structure of the same crystal~\cite{Bianchi2010}. This DP binding energy is typical for naturally grown Bi$_2$Se$_3$ crystals, which are known to be electron doped due to the presence of Se vacancies observable in STM images (Fig.~\ref{STM}(a))~\cite{Urazhdin2002,arxivpaperAlpichshev}. Upon deposition of Fe (Fig.~\ref{STM}(c)), the substrate spectral intensity reveals a minimum that is shifted downwards with respect to the uncovered surface by about 50~meV. Since, for the uncovered surface, this minimum coincides with the DP, we conclude that Fe probably acts as a donor and the DP is accordingly shifted downwards as seen by recent ARPES measurements~\cite{arxivpaperScholz}. Most notably, a resonance can be seen in spectra taken on the Fe atoms. It appears at voltages slightly above the DP and has differing intensities, peak positions, and linewidths depending on the binding site. Substrate corrected spectra (inset) prove that the resonance has a considerably stronger intensity and is shifted closer to the DP for the type B Fe atoms. Similar resonances have been observed on the Se vacancies~\cite{Urazhdin2002,arxivpaperAlpichshev} and were attributed to Coulomb scattering of the topological surface state by the impurity~\cite{Biswas2010,arxivpaperAlpichshev}.

\begin{figure}
\includegraphics[width=\columnwidth]{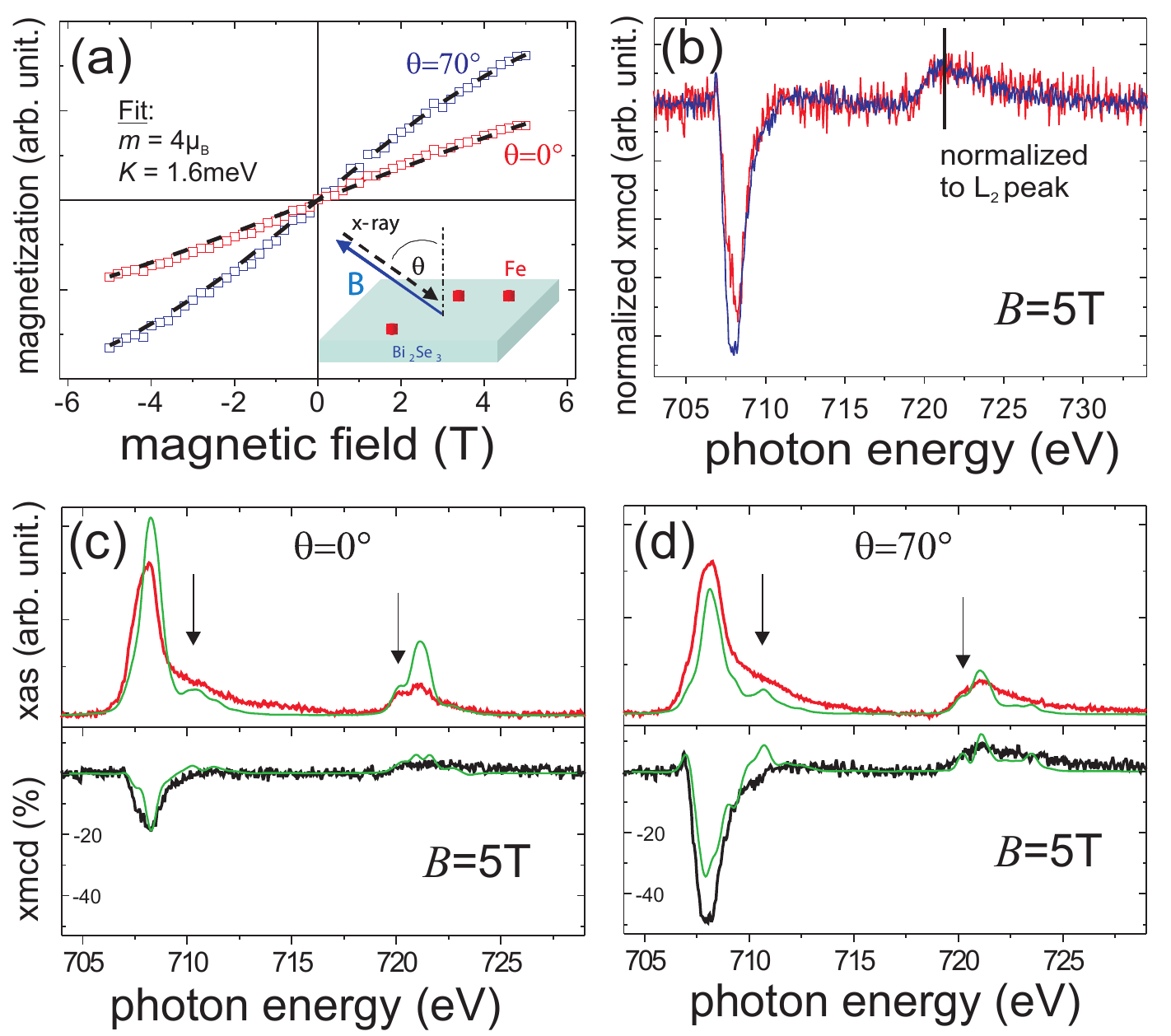}
\caption{\label{x-ray} X-ray study of Fe impurities (coverage $1\%$ML) on Bi$_2$Se$_3$ at $T=10$~K: (a) Magnetization curves $M$($B$) for directions $\theta = 0^{\circ}, 70^{\circ}$ in red and blue, respectively. The dotted lines represent thermodynamic fits with a moment $m=4$~$\mu_{B}$ and $K=1.6$~meV (see text). (b) XMCD signals in a field of $B=5$~T normalized to the XMCD peak intensity at the $L_{2}$-edge. (c) and (d): XAS (red, upper panel) and XMCD (black, lower panel) measured in a field of $B=5$~T for directions $\theta = 0^{\circ}$ (c) and $70^{\circ}$ (d). Green spectra refer to CF calculations with $D_{\sigma}=0.19$~eV and $D_{\tau}=0.05$~eV.}
\end{figure}

For a magnetic characterization, x-ray absorption spectra (XAS) at the Fe $L_{3,2}$ absorption edges (between $690$~eV and $740$~eV) were recorded in the surface sensitive total electron yield mode for different angles $\theta$ between the Bi$_2$Se$_3$ surface normal and the x-ray beam direction. Magnetic fields $B$ were applied collinear to the beam as shown in the inset of Fig.~\ref{x-ray}(a). The experimental timescales for measuring spectra are on the order of $100$~s. Before and after the magnetic characterization, the samples were carefully checked for oxygen contaminations using the XAS at the O-1$s$ absorption edge.

Typical non-dichroic XAS defined as the average of XAS for right ($\sigma^{+}$) and left ($\sigma^{-}$) circular polarization are shown after Bi$_2$Se$_3$ background subtraction in the upper panel of Fig.~\ref{x-ray}(c) and (d). When $B$ is applied, a pronounced XMCD signal defined as $(\sigma^{+} - \sigma^{-})$ is observed, which reflects the spin and orbital polarization in the Fe~3$d$ states. As a first approximation the XMCD signal is proportional to the average Fe magnetization $M$ when it is normalized to the $L_{3}$ peak intensity in the XAS signal as shown in the lower panels of Fig.~\ref{x-ray}(c) and (d) for fields of $B=5$~T. More precisely, ${\text{XMCD/XAS}}|_{L_3} \propto \vec{\Sigma} \cdot \vec{M}$, i.e. the XMCD signal normalized to the total XAS intensity is proportional to the projection of the averaged Fe moments $<\vec{m}>$ on the photon propagation direction $\vec{\Sigma}$. Comparing the lower panels of Fig.~\ref{x-ray}(c) and (d) we thus conclude that Fe atoms prefer to be magnetized in the $\theta = 70^{\circ}$ direction as compared to the out-of-plane case where $\theta = 0^{\circ}$.

This is supported by the significant anisotropy in the orbital moment $m_L$ visible in Fig.~\ref{x-ray}(b): the XMCD
spectra are normalized to their peak intensity at the $L_{2}$ edge ($721.0$~eV) and smaller intensities in the polar
geometry at the $L_{3}$ edge ($708.0$~eV) indicate considerably reduced orbital moments $m_L$ with respect to the direction $\theta
= 70^{\circ}$. By measuring the normalized XMCD intensity at the $ L_{3}$ edge  for different magnetic fields
it is possible to obtain magnetization curves $M$($B$). The curves $M$($B$) in Fig.~\ref{x-ray}(a) at $\theta =
0^{\circ}, 70^{\circ}$ show smaller values along $\theta = 0^{\circ}$ again proving a preferential orientation of
the magnetization in the (111) plane of Bi$_2$Se$_3$.

In the following we want to discuss the experimentally observed magnetic anisotropy of the Fe adatoms. 
Let us first look at the XAS lineshape in Fig.~\ref{x-ray}(c) and (d),
which reflects the electronic configuration and possible crystal field (CF) effects from the Bi$_2$Se$_3$ surface. A pronounced double-peak structure is observed at the $L_{2}$ edge as
well as a shoulder at the high-energy side of the $L_{3}$ edge (see arrows). CF multiplet calculations, including electron repulsion and spin-orbit coupling~\cite{Stepanow2010} can reproduce the main features assuming a purely trigonal CF of D$_{3h}$
symmetry: the green curves represent spectra derived for Fe $d^6$ impurities in a CF defined by the parameters
$D_{\sigma}=0.19$~eV and $D_{\tau}=0.05$~eV.
The double-peak structure in the XAS of the $L_{2}$ edge at $\theta = 0^{\circ}$ as well as the shoulder at the $L_{3}$ edge for both $\theta = 0^{\circ}$ and $70^{\circ}$ are reproduced.
The measured spectra are broadened with respect to the model, which points towards a significant coupling of $d$-states to delocalized bands.
From our CF calculations also considering Boltzmann statistics at $T=10$~K, we derive anisotropic orbital moments with $\Delta m_L=0.13$~$\mu_{B}$ in the two measuring geometries supporting the experimentally observed quenching of the orbital moments in the $\theta=0^{\circ}$ direction under the influence of the in-plane CFs. Assuming a moment of $|\vec{m}|=4$~$\mu_{B}$ we can fit the experimental magnetization curves in Fig.~\ref{x-ray}(a) with a thermodynamic model which includes the Zeeman term and an anisotropy term $E_{\rm A}=K\cdot\left(\vec{z} \cdot \vec{m}/\left|m\right|\right)^2$ where $K$ is the magnetic anisotropy per Fe atom and $\vec{z}$ the normal of the (111) plane. The fit leads to a significant in-plane magnetic anisotropy of $K=+1.6$~meV per atom.

\begin{figure}
\includegraphics[width=\columnwidth]{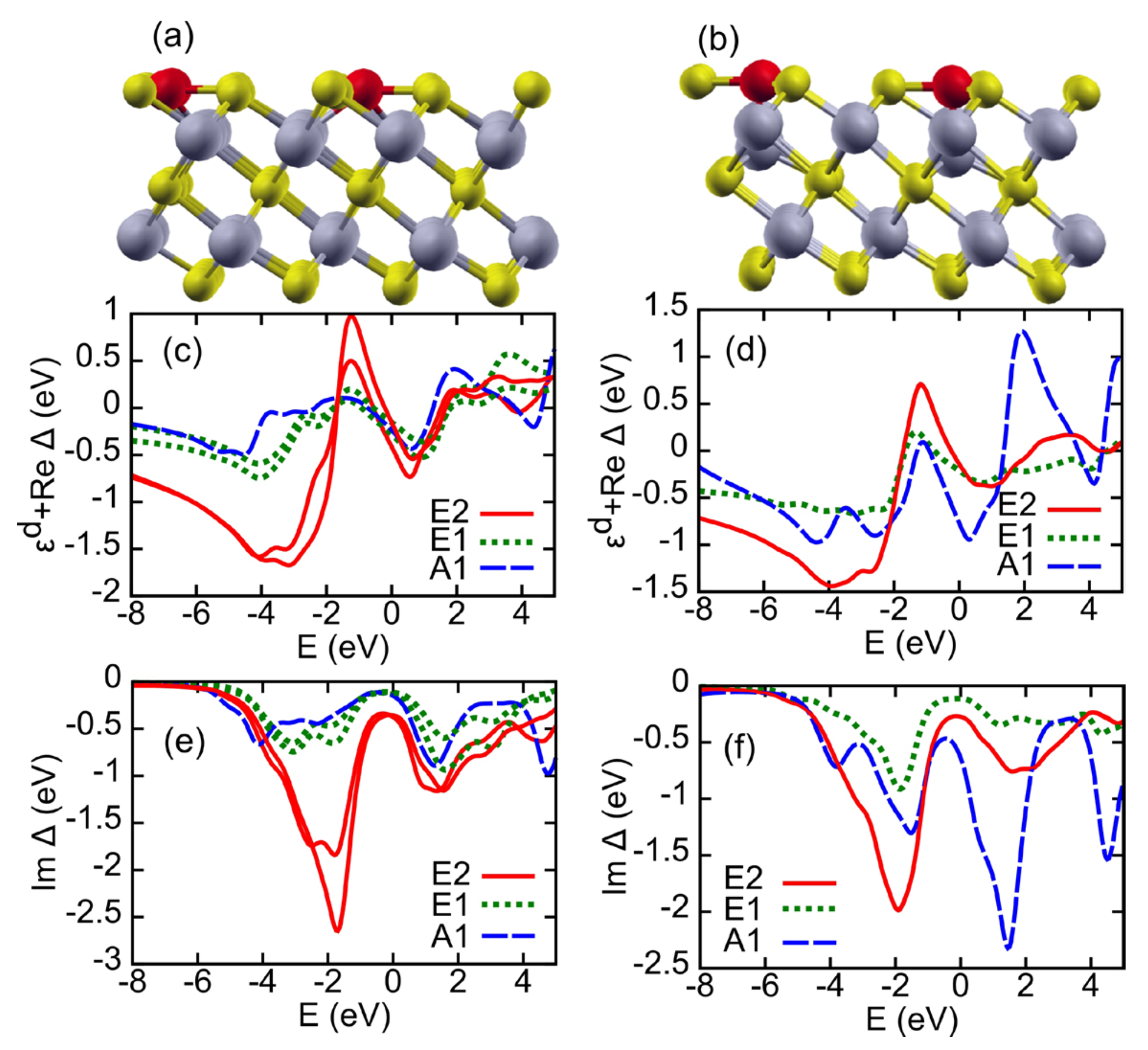}


\caption{\label{DFT} (a), (b): Adsorption geometries of Fe (red) on Bi$_2$Se$_3$ (Bi: gray, Se: yellow). Fe can adsorb at an fcc (a) or an hcp hollow site (b). The real and imaginary parts of the hybridization functions ($\Re\Delta_{ii}(E)+\epsilon^d_{ii}$ and  $\Im\Delta_{ii}(E)$) are given in (c,d) and (e,f), respectively, and  quantify the interaction of the Fe $3d$-electrons with their environment. $\Delta_{ii}(E)$ is shown for all Fe-$3d$ orbitals: E2 ($d_{xy},d_{x^2-y^2}$), E1 ($d_{xz},d_{yz}$), and A1 ($d_{z^2}$).}
\end{figure}

To understand the physical mechanisms behind the experimentally observed magnetic properties of the system we resort to {\it ab initio} calculations and derive a quantum impurity model describing the Fe adatoms on the Bi$_2$Se$_3$ surface.
We performed density functional (DFT) calculations of Fe adatoms on the Bi$_2$Se$_3$ surface which yield the relaxed adsorption geometries, the magnetic moments, and the hybridization functions~\cite{PAW_DMFT,DFT++} \footnote{Here, we employ 6 quintuple layers thick $2\times 2$ Bi$_2$Se$_3$ surface supercells containing one Fe adatom and we use the Vienna Ab Initio Simulation Package (VASP)~\cite{Kresse:PP_VASP} with the generalized gradient approximation (GGA)~\cite{PBE} to the exchange correlation potential and the projector augmented wave (PAW)~\cite{Bloechl:PAW1994,Kresse:PAW_VASP} basis sets to solve the resulting Kohn-Sham system. We relaxed the positions of the Fe atoms and of the atoms in the uppermost Bi$_2$Se$_3$ quintuple layer until the forces were below $0.02$~eV\AA$^{-1}$.}.

Our calculations reveal two possible adsorption sites for an Fe adatom at an fcc hollow (f-) or an hcp hollow (h-) site [Fig.~\ref{DFT} (a,b)]. While both adsorption geometries are stable, we find that the f-site
adsorption yields $0.07$~eV lower total energy than the h-site. Independent of the adsorption site, GGA predicts a Fe $3d$-occupancy of $N_d=6.3$.
At both sites, we find a strong relaxation of the adatom into the surface with equilibrium positions of the Fe adatom virtually at the same height ($\pm 0.1$~\AA ) as the surrounding Se atoms. This is well in line with the STM topography (Fig.~\ref{STM}a) and the observed broadening in the x-ray absorption spectra. It has important consequences for the electronic and magnetic properties of the Fe adatoms.

The energy dependent coupling strength of the impurity $3d$-orbitals and the surrounding crystal is quantified by the hybridization functions $\Delta_{ij}(E)$ where $i,j=(m,\sigma)$ denote combined orbital and spin quantum numbers of the Fe $3d$ electrons and $E$ is the energy. These hybridization functions can be interpreted as energy dependent complex valued potentials acting on the Fe $3d$ orbitals and fully characterize their interaction with the substrate as regards local observables (see Ref.~\cite{Wehling2011} and the supplement for further details).
For Fe in the h-site position (Fig. \ref{DFT}(f)), we find the most pronounced peaks in $\Im \Delta_{ii}(E)$ for the in-plane $d$-orbitals ($d_{xy},d_{x^2-y^2}$ spanning the subspace of orbital angular momentum quantum number $|l_z|=2$) at $E\approx -2$~eV and for the out-of-plane oriented $d_{z^2}$ orbital above the Fermi level. For Fe at the energetically most favorable
f-site (Fig. \ref{DFT}(e)), $\Im \Delta_{ii}(E)$ is clearly dominated by peaks at $E\approx-2$~eV in the $d_{xy}$ and $d_{x^2-y^2}$ orbitals. The hybridization of the $d_{z^2}$ orbital is smaller than at the h-site, because there is no Bi atom directly beneath the Fe atom. Consequently, at both adsorption sites, the in-plane oriented $d_{xy}$ and $d_{x^2-y^2}$-orbitals hybridize strongly with electronic states from the surrounding crystal. The relaxation of the Fe adatoms into the Bi$_2$Se$_3$ surface facilitates the strength of this particular hybridization channel.

$\Re \Delta_{ii}(E)$ can be interpreted as the energy dependent crystal field, which controls the occupation of the impurity orbitals. Fe has $N_d\sim 6$ $d$ electrons, five of which occupy the spin-up $d$ orbitals. The remaining spin-down $d$ electron determines the magnetic easy axis (see supplement for details). If it occupies an orbital with $|l_z|>0$, one obtains an out-of-plane easy axis, whereas occupation of the $d_{z^2}$ orbital ($l_z=0$) leads to an in-plane easy axis. The resonance in the hybridization function of the in-plane $d$-orbitals at $-2$~eV indicates a resonance of the Bi$_2$Se$_3$ host at this energy which strongly couples to the Fe ($d_{xy},d_{x^2-y^2}$) orbitals. As a consequence, these Fe orbitals are pushed upwards within the energy region $-2\,\rm{eV}\lesssim E \lesssim 0\,\rm{eV}$ (c.f. Fig.~\ref{DFT}(c) and (d)) due to level repulsion between the resonance of the Bi$_2$Se$_3$ host and the Fe ($d_{xy},d_{x^2-y^2}$) orbitals. Therefore, the Fe ($d_{xy},d_{x^2-y^2}$) orbitals become depopulated, which facilitates in-plane orbital momenta and an in-plane magnetic easy axis (see supplement for a detailed analysis in terms of an Anderson impurity model).

The crystal fields obtained from fitting the spectra of CF multiplet calculations to the XAS/XMCD experiments were $D_{\sigma}=0.19$~eV and $D_{\tau}=0.05$~eV. These values agree qualitatively with the dynamical crystal fields concerning the emptied orbitals as seen by the red curves at energies between $E=-2$~eV and $-1$~eV in Fig.~\ref{DFT}(c) and (d). The fitted spectra are thus in line with the picture of dynamic hybridization effects leading to a depopulation of the Fe ($d_{xy},d_{x^2-y^2}$) orbitals and a resulting in-plane magnetic easy axis.





In summary, utilizing a combination of experiment and theory, we demonstrate that the favorable orientation of the magnetic
moment of Fe impurities absorbed on the Bi$_2$Se$_3$ surface is in-plane with a
considerable anisotropy but no hysteretic behaviour, i.e. magnetic ordering. This has important consequences for the stability of the topological Bi$_2$Se$_3$ surface state against perturbations by \textit{magnetic} Fe impurities, which break time-reversal symmetry: On average, the exchange field by the Fe impurities acts like an in-plane oriented magnetic field $\vec{B}$ on the spins of the surface state Dirac fermions: $H_D=v_f(\vec{\sigma}\cdot\vec p)+g\mu_B\vec B\cdot\vec{\sigma}$ \footnote{Here $v_{f}$
is the Fermi velocity, $g$ the effective Land{\'e} factor of the surface state electrons, $\vec\sigma$ are Pauli matrices acting on the spin of the surface state electrons and $p$ is their momentum.}. From this Dirac Hamiltonian it becomes immediately clear, that an in-plane exchange field $\vec{B}$ on average merely shifts the DP to $\vec p_D=-g\mu_B \vec{B}/v_f$ but it does not open a gap, despite of the breaking of time-reversal symmetry. This qualitatively agrees with new photoemission spectra which do not reveal a gap at the DP~\cite{arxivpaperScholz,Hofmann2011} as previously observed~\cite{Wray2011}.
Furthermore, we show that the favored orientation of the moment can be well understood from the interplay of local atomic physics with dynamic hybridization effects. This illustrates the necessity to account for both, local Coulomb interaction and dynamical hybridization effects, when considering {\it ab initio} approaches for these systems.

\textbf{Acknowledgments}\\
Financial support from the ERC Advanced Grant ``FURORE'', by the Deutsche Forschungsgemeinschaft via the SFB668, by the city of Hamburg via the cluster of excellence ``Nanospintronics'', by the Danish Council for Independent Research, the Danish National Research Foundation, as well as computer time at HLRN (Germany) are gratefully acknowledged. We thank J. Kolorenc for providing us his exact diagonalization code, and S. V. Eremeev and E. V. Chulkov for fruitful discussions.





\end{document}